\pdfoutput=1
\documentclass[%
 reprint,
 twocolumn,
 amsmath,amssymb,
 aps,
]{revtex4-1}

\usepackage{graphicx}
\usepackage{dcolumn}
\usepackage{bm}
\usepackage{tikz}
\usepackage{tikz-feynman}
\usepackage{relsize}
\usepackage{physics}

\tikzset{
    ->-/.style={decoration={
  markings,
  mark=at position .6 with {\arrow{>}}},postaction={decorate}},
    -<-/.style={decoration={
  markings,
  mark=at position .5 with {\arrow{<}}},postaction={decorate}},
}

\newlength{\mylen}
\setbox1=\hbox{$\bullet$}\setbox2=\hbox{\tiny$\bullet$}
\usetikzlibrary{positioning}
\tikzset{main node/.style={circle,line width=0.5mm,fill=blue!20,draw,pattern=crosshatch dots,pattern color=black!50!white,minimum size=1.2cm,inner sep=0pt},
            }
\tikzset{main node2/.style={circle,line width=0.5mm,fill=blue!40,draw,pattern=crosshatch dots,pattern color=black!50!white,minimum size=3cm,inner sep=0pt},
            }
            
\tikzset{main node3/.style={circle,line width=0.5mm,fill=blue!40,draw,pattern=crosshatch dots,pattern color=black!50!white,minimum size=2.3cm,inner sep=0pt},}

\usepackage{hyperref}

\usepackage{xcolor}

\usepackage{microtype}
\usepackage{breakurl}

\newcommand*{\colorboxed}{}
\def\colorboxed#1#{%
  \colorboxedAux{#1}%
}
\newcommand*{\colorboxedAux}[3]{%
  \begingroup
    \colorlet{cb@saved}{.}%
    \color#1{#2}%
    \boxed{%
      \color{cb@saved}%
      #3%
    }%
  \endgroup
}

\chardef\MyArticleWithColor=\pdfcolorstackinit page direct{0 g}

\chardef\MyArticleWithColor=\pdfcolorstackinit page direct{0 g}

\chardef\MyArticleWithColor=\pdfcolorstackinit page direct{0 g}

\usepackage{dsfont}
\begin{document}

\preprint{APS/123-QED}

\title{Exposing the threshold structure of loop integrals}

\author{Zeno Capatti}
 \email{zeno.ca@gmail.com}
\affiliation{%
 Institute for Theoretical Physics, ETH Z\"urich, \\ Wolfgang-Pauli-Str. 27, 8093, Z\"urich  
}%

\date{\today}

\begin{abstract}



The understanding of the physical laws determining the infrared behaviour of amplitudes is a longstanding and topical problem. In this paper, we show that energy conservation alone implies strong constraints on the threshold singularity structure of Feynman diagrams. In particular, we show that it implies a representation of loop integrals in terms of Fourier transforms of non-simplicial convex cones. We then engineer a triangulation that has a direct diagrammatic interpretation in terms of a straightforward edge-contraction operation. We use it to develop an algorithmic procedure that performs the Fourier integrations in closed form, yielding the novel Cross-Free Family three-dimensional representation of loop integrals. Contrary to the TOPT and LTD representations, its singularity structure is entirely and elegantly expressed in terms of the graph-theoretic notions of connectedness and crossing. These results can be used to classify infrared-finite scattering theories, numerically evaluate loop integrals and to simplify threshold regularisation procedures. 
\end{abstract}

\maketitle


\section{\label{sec:introduction}Introduction}

The study of the threshold singularity structure of Feynman diagrams and scattering amplitudes is an enduring effort that has, throughout the years, lead to incredible developments that have shed light on the infrared physics of quantum scattering phenomena (see~\cite{Agarwal:2021ais} for a review and~\cite{Cutkosky:1960sp,PhysRev.52.54,Jauch,YENNIE1961379,Kinoshita:1962ur,Lee:1964is,PhysRevD.17.2773,PhysRevD.17.2789,Frye:2018xjj,Agarwal:2021ais,Hannesdottir:2022bmo,Hannesdottir:2019rqq,Magnea:2021fvy,Ma:2019hjq,Erdogan:2014gha,Akhoury:2011kq,Binoth:2000ps,Libby:1978bx,Collins:2020euz,Mizera:2021icv,Erdogan:2013bga,Heissenberg:2021tzo,Caron-Huot:2017zfo,Agarwal:2020nyc,Catani:1996vz,Catani:1998bh,Dixon:2008gr,Becher:2009cu,Becher:2009qa,Gardi:2009zv,Dixon:2009ur,Ahrens:2012qz,Naculich:2013xa,Becher:2009kw,Ferroglia:2009ii,Gardi:2010rn,Aguilera-Verdugo:2019kbz,Doria:1980ak,Henn:2016jdu,Arkani-Hamed:2020gyp,Bern:1999ry,Dennen:2016mdk,Caola:2020xup,PARISI1990455,DelDuca:1990gz,Laenen:2008gt,Contopanagos:1992fm,Forde:2003jt,1965NCimS..38..438C} for a historical selection of works on the topic).

Three-dimensional representations of Feynman diagrams, obtained via Time-Ordered Perturbation Theory (TOPT) (see~\cite{Sterman:1993hfp,tasi_sterman} for a review), via Flow-Oriented Perturbation Theory (FOPT)~\cite{Borinsky:2022msp} or through the Loop-Tree Duality (LTD) formalism~\cite{Catani_2008,Bierenbaum_2010,Runkel_2019,Runkel_2020,Capatti_2019,Tomboulis:2017rvd}, are famously apt to performing a systematic singularity analysis of Feynman integrals. Indeed, the interplay of energy conservation and residue theorem involved in their derivation makes their singularities directly interpretable in terms of cuts~\cite{Cutkosky:1960sp}. This in turn allows to leverage a host of graph-theoretical knowledge to perform a diagrammatic study of the structure of physical thresholds. 
Even then, these three-dimensional representations are plagued by spurious divergences, corresponding to cuts that divide the graph in more than two connected components (TOPT) or cuts containing particles that have both positive and negative on-shell energy (LTD). Even improved LTD formalisms~\cite{capatti2020manifestly,Aguilera_Verdugo_2020,Capatti:2020xjc,Aguilera_Verdugo_2021,Sborlini_2021,Kromin:2022txz,bobadilla2021lotty,Ramirez-Uribe:2020hes,Ramirez-Uribe:2022sja} that remove spurious singularities by relying on algebraic manipulations of the integrand or direct ansatzes are either inadequate for wider generalisations or lack first-principle justification.

In this paper, we derive the precise relationship between the singularity structure of a Feynman integral and the graph-theoretic notions of crossing and connectedness~\cite{Abreu:2014cla,Bloch:2015efx,Arkani-Hamed:2017fdk,Abreu:2017ptx,Capatti:2020xjc,bourjaily2020sequential,Berghoff:2020bug,Kreimer:2020mwn,Kreimer:2021jni,Benincasa:2020aoj,Sborlini_2021,Hannesdottir}, by explicitly constructing a three-dimensional representation of loop integrals that manifests such properties.

We do so by exploiting methods that pertain to a recently growing branch of literature that applies methods of convex geometry, and more precisely the geometry of polytopes~\cite{Hodges:2009hk,Arkani-Hamed:2012zlh,Arkani-Hamed:2013jha,Arkani-Hamed:2014dca,Arkani-Hamed:2017fdk,Arkani-Hamed:2022cqe, Benincasa:2020aoj, Benincasa:2021qcb, Borinsky:2020rqs,https://doi.org/10.48550/arxiv.1512.06409,Binoth:2000ps,Kaneko_2010,Pak_2011,Tourkine_2017,https://doi.org/10.48550/arxiv.1408.3410,Kreimer:2020mwn,Kreimer:2021jni} and their Fourier transform~\cite{Borinsky:2022msp}, to problems in high-energy physics. In particular, we will relate Feynman diagrams to Fourier transforms of (non-simplicial) convex cones. We will then perform the Fourier integration analytically using an identity with a digrammatic interpretation in terms of a straightforward edge-contraction procedure. 

We provide an algorithm based on the recursive application of edge contraction, resulting in the surprisingly compact and elegant \textit{Cross-Free Family} (CFF) \textit{representation} of loop integrals. Each denominator is identified with a cross-free family of connected subgraphs of the starting graph. Such representation can be cast in factorised form and is especially suited for the numerical evaluation of loop integrals. We provide in the mathematica package \texttt{cLTD.m}~\footnote{ \href{http://www.github.com/apelloni/cLTD}{github.com\slash apelloni\slash cLTD}} a generic implementation of the algorithm presented in this paper.

 We finally describe the factorisation properties of the Fourier transform which, as a result, provide a necessary condition for the non-integrability of the intersection of any number of threshold singularities for QCD diagrams. The CFF representation, together with the factorisation argument, de facto provide a classification of the threshold singularity structure of Feynman diagrams in terms of crossing and connectedness, a fundamental step in the long-standing quest for the understanding of the infrared behaviour of Feynman diagrams and amplitudes.


\section{Energy conservation}

Consider a bridgeless digraph $G=(\mathcal{V},\mathcal{E})$ with underlying undirected graph $G_u$, given as a tuple of a set of vertices $\mathcal{V}$ and a set of ordered pairs of vertices $\mathcal{E}$. To each edge $e\in\mathcal{E}$ of the graph is associated a weight $x_e\in\mathbb{R}$, collected in a vector $\mathbf{x}=(x_e)_{e\in \mathcal{E}}\in \mathbb{R}^{|\mathcal{E}|}$.  The standard scalar product of two weight vectors $\mathbf{x}$ and $\mathbf{y}$ is defined as $\mathbf{x}\cdot \mathbf{y}=\sum_{e\in \mathcal{E}}x_e y_e$. The component-wise multiplication of two vectors $\boldsymbol{\sigma}$ and $\mathbf{x}$ is denoted as $\mathbf{x}\odot\boldsymbol{\sigma}=(\sigma_e x_e)_{e\in \mathcal{E}}$. Given any subset of the edges $\mathcal{E}'\subseteq \mathcal{E}$, its characteristic vector $\mathbf{1}^{\mathcal{E}'}$ has components $\mathbf{1}^{\mathcal{E}'}_e=1$ if $e\in \mathcal{E}'$ and $\mathbf{1}^{\mathcal{E}'}_e=0$ otherwise. 
The positive (negative) boundary of a subset $S\subset\mathcal{V}$ (also called a \emph{cut} $S$), is defined as
\begin{align}
    &\delta^{+}(S)=\{e=\{v,v'\}\,|\, v\in S, v'\notin S\}, \\
    &\delta^{-}(S)=\{e=\{v',v\}\,|\, v\in S, v'\notin S\},
\end{align} 
and $\delta(S)=\delta^{+}(S)\cup \delta^{-}(S)$. Consider the space of weight vectors satisfying energy-conservation constraints, $\mathcal{Q}^0_G\subset\mathbb{R}^{|\mathcal{E}|}$. An element $\mathbf{q}^0_G\in\mathcal{Q}^0_G$ satisfies energy conservation at each vertex:
\begin{equation}
      \mathbf{q}^0_G\cdot(\mathbf{1}^{\delta^+(v)}-\mathbf{1}^{\delta^-(v)})=p_v^0, \quad \forall v\in\mathcal{V},
\end{equation} 
for fixed vertex weights $\{p_v^0\}_{v\in\mathcal{V}}$ with $\sum_{v\in\mathcal{V}}p_v^0=0$. In Feynman diagrams, $p_v^0$ is the external momentum that injects into the vertex $v$, which may be chosen to be vanishing. Elements of $\mathcal{Q}^0_G$ can be written in terms of elements of a cycle basis $\mathcal{C}$ of the graph as
\begin{equation}
\label{eq:loop_decomposition}
    \mathbf{q}^{0}_G(\{k_c^0\}_{c\in\mathcal{C}}, \{p_v^0\}_{v\in\mathcal{V}})=\sum_{
    c\in \mathcal{C} 
    } \mathbf{s}_{c}^{G} k_c^0 + \mathbf{p}^0_G(\{p_v^0\}_{v\in\mathcal{V}}), 
\end{equation}
with $L=|\mathcal{C}|=|\mathcal{E}|-|\mathcal{V}|+1$, and
\begin{equation}
\label{eq:external_momentum}
    \mathbf{p}^0_G=\sum_{v\in\mathcal{V}} \mathbf{r}_{v}^G p_v^0.
\end{equation}
$\mathbf{s}^G_{c}$ is constructed in the following way: first, $(\mathbf{s}^G_{c})_e=s^G_{ce}=0$ if and only if $e$ does not belong to $c$. Second, $s^G_{ce}=1$ if $e$ belongs to the cycle $c$ and the cycle $c$ is oriented in $G$. Finally, $s^{G'}_{ce}=-s^{G}_{ce}$, if the orientation for the arc $e$ in $G$ is swapped, yielding a graph $G'$. The cycle vector so constructed is ambiguous up to a sign, which can be chosen arbitrarily. Analogously, we have $r^{G'}_{ve}=-r^{G}_{ve}$ (for the purposes of this paper, this is all we need to know about $\mathbf{r}_v^G$).  

\section{\label{sec:cone_fourier}Feynman diagrams as Fourier transforms of convex cones}

A bridgeless Feynman diagram $I_{G_\text{u}}(\mathbf{E},\{p_v^0\}_{v\in\mathcal{V}})$, stripped of its spatial loop integrations, reads
\begin{equation}
\label{eq:Feynman_diagram}
    I_{G_\text{u}}=\int \left[\prod_{c\in\mathcal{C}} \frac{\mathrm{d} k_c^0}{2\pi}\right] \frac{\mathcal{N}_G(\mathbf{q}_G^0)}{\prod_{e\in \mathcal{E}}((q_e^0)^2-E_e^2+i\varepsilon)},
\end{equation}
where $q_e^0=(\mathbf{q}^0_G)_e$ is the linear function of the loop momenta given in eq.~\eqref{eq:loop_decomposition}, for an arbitrarily chosen digraph $G$ with underlying graph $G_u$, and $\mathcal{N}_G$ is a polynomial numerator. $I_{G_\text{u}}$ is parametric in $\mathbf{E}$, which for applications in QFT is the vector of on-shell energies assigned to the edges of the graph, $E_e=\sqrt{|\vec{q}_e|^2+m_e^2}$. We introduce an auxiliary integration $1=\int \mathrm{d}x_e\delta(x_e-q_e^0)$ for each edge and use the Fourier representation of the Dirac delta function, which yields
\begin{equation}
\label{eq:Feynman_diagram_delta}
    I_{G_u}=\int \prod_{c\in\mathcal{C}} \frac{\mathrm{d} k_c^0}{2\pi} \int \mathrm{d}\mathbf{x} \mathrm{d}\boldsymbol{\tau} \, \frac{\mathcal{N}_G(\mathbf{x})e^{i\boldsymbol{\tau}\cdot(\mathbf{x}-\mathbf{q}^0_G)}}{\prod_{e\in \mathcal{E}}2\pi (x_e^2-E_e^2+i\varepsilon)} .
\end{equation}
This step requires one auxiliary variable for each edge (as in the FOPT derivation~\cite{Borinsky:2022msp}) and not for each vertex (as in TOPT). 
Residue integration in the variables $x_e$ appearing in the propagators of eq.~\eqref{eq:Feynman_diagram_delta} is trivial, as they are not subject to energy conservation. Each propagator contributes with two terms, associated to the sign of $\tau_e$ and corresponding to the poles $x_e=\pm E_e$. Thus, we express $I_{G_u}$ as a sum over $2^{|\mathcal{E}|}$ contributions
\begin{equation}
        I_{G_u}=\sum_{\boldsymbol{\sigma}\in \{\pm 1\}^{|\mathcal{E}|}} \frac{\mathcal{N}_G(\boldsymbol{\sigma}\odot\mathbf{E})}{\prod_{e\in \mathcal{E}}2iE_e}\hat{\mathds{1}}_{\boldsymbol{\sigma}}(\mathbf{E},\{p_v^0\}_{v\in\mathcal{V}}),
\end{equation}
where we introduced, with some fore-shadowing, the function $\hat{\mathds{1}}_{\boldsymbol{\sigma}}(\mathbf{E},\{p_v^0\}_{v\in\mathcal{V}})$ defined as
\begin{align}
    \hat{\mathds{1}}_{\boldsymbol{\sigma}}=\int& \prod_{c\in\mathcal{C}} \frac{\mathrm{d} k_c^0}{2\pi}  \int \prod_{e\in \mathcal{E}}\mathrm{d} \tau_e e^{i\tau_e(\sigma_eE_e-q_e^0)}\Theta(-\sigma_e\tau_e) .
\end{align}
Note that, in order for the residue at infinity to vanish, $\mathcal{N}$ must be a polynomial in which the energy arguments appear with power at most one~\footnote{The same condition holds for the derivation of TOPT presented in~\cite{tasi_sterman,Sterman:1993hfp}}. We observe that any diagram can itself be algebraically decomposed into diagrams satisfying this numerator property. Changing integration variables from $\tau_e$ to $-\sigma_e\tau_e$ gives
\begin{equation}
   \hat{\mathds{1}}_{\boldsymbol{\sigma}}=\int \prod_{c\in\mathcal{C}} \frac{\mathrm{d} k_c^0}{2\pi} \int_{\mathbb{R}_+^{|\mathcal{E}|}}\mathrm{d} \boldsymbol{\tau}e^{-i\boldsymbol{\tau}\cdot (\mathbf{E}-\boldsymbol{\sigma}\odot\mathbf{q}^0_G )}.
\end{equation}
 Plugging eq.~\eqref{eq:loop_decomposition} in $\boldsymbol{\tau}\cdot (\boldsymbol{\sigma}\odot\mathbf{q}^0)$ we factorise integration on $k_c^0$, which is then performed by using the Fourier representation of the Dirac delta function, but in reverse, 
\begin{equation}
\label{eq:fourier_transform_delta}
    \hat{\mathds{1}}_{\boldsymbol{\sigma}}=\int_{\mathbb{R}_+^{|\mathcal{E}|}}\mathrm{d}\boldsymbol{\tau}  e^{-i\boldsymbol{\tau}\cdot(\mathbf{E}-\boldsymbol{\sigma}\odot\mathbf{p}^0_G)}\prod_{c\in\mathcal{C}}\delta\left(\boldsymbol{\tau}\cdot\left(\boldsymbol{\sigma}\odot \mathbf{s}_c^G \, \right)\right) .
\end{equation}
The Dirac delta functions enforce the integration domain to be the intersection of $|\mathcal{C}|$ hyperplanes with the positive orthant $\mathbb{R}_+^{|\mathcal{E}|}$. 
Observe that, by definition, $\boldsymbol{\sigma}\odot\mathbf{s}_c^G=\mathbf{s}_c^{G'}$, $\boldsymbol{\sigma}\odot\mathbf{p}^0_G=\mathbf{p}^0_{G'}$ and $\mathcal{N}_{G'}(\mathbf{E})=\mathcal{N}_{G}(\boldsymbol{\sigma}\odot\mathbf{E})$ where $G'$ is the graph obtained from $G$ by swapping the orientation of the edge $e$ if $\sigma_e=-1$. We can thus substitute the sum over $\boldsymbol{\sigma}$ with a sum over digraphs. Let us also observe that, if the graph has an oriented cycle $c$, then we must have $\boldsymbol{\tau}\cdot \mathbf{s}_c^{G'}=\sum_{e\in c}\tau_e=0$ in virtue of the Dirac deltas of eq.~\eqref{eq:fourier_transform_delta}. Since $\tau_e>0$ for any $e\in \mathcal{E}$, this implies that the integration domain for the Fourier transform is empty. Consequently, the orientations $G$ that contribute with non-vanishing Fourier transform correspond to directed acyclic graphs. In summary, let $\text{dag}(G_u)$ be the set of all acyclic digraphs with underlying graph $G_u$, so that 
\begin{equation}
\label{eq:sum_over_digraphs}
    I_{G_u}=\sum_{G\in\text{dag}(G_u)} \frac{\mathcal{N}_G(\mathbf{E})}{\prod_{e\in \mathcal{E}}2iE_e}\hat{\mathds{1}}_{\mathcal{K}_G},
\end{equation}
with
\begin{equation}
  \hat{\mathds{1}}_{\mathcal{K}_{G}}=\int_{\mathcal{K}_{G}}\mathrm{d}\boldsymbol{\tau}  e^{-i\boldsymbol{\tau}\cdot(\mathbf{E}-\mathbf{p}_{G}^0)},
\end{equation}
where $\hat{\mathds{1}}_{\mathcal{K}_{G}}$ is the Fourier transform of the characteristic function $\mathds{1}_{\mathcal{K}_G}$ of the cone $\mathcal{K}_G$, defined as
\begin{align}
    \mathcal{K}_G=\left\{\boldsymbol{\tau}\in\mathbb{R}_+^{|\mathcal{E}|} \ \Big| \ \boldsymbol{\tau}\cdot \mathbf{s}_c^G=0, \ \forall c\in\mathcal{C}\right\}.
\end{align}
Only acyclic digraphs contribute to the sum. This is analogous to TOPT, expressed as a sum over vertex orderings, or \emph{topological orderings} in technical jargon. To each topological ordering corresponds a unique acyclic orientation (but not viceversa). The relevance of acyclic graphs has also been recognised in the context of causal representations~\cite{Sborlini_2021}. In FOPT~\cite{Borinsky:2022msp}, a dual phenomenon happens: in that case, the three-dimensional representation is expressed as a sum over strongly-connected digraphs.

\section{\label{sec:level2} Triangulations and the edge-contraction operation}

Before engineering a  triangulation of the convex cones $\mathcal{K}_G$, let us discuss the case in which $G$ is a multigraph, i.e. it has multiple edges connecting the same two vertices. Let ${e_1,...,e_n}$ be a set of edges that connect the same two vertices, $v$ and $v'$. Let $G'$ be the graph in which all such edges have been substituted by a unique edge $e$. Then
\begin{equation}
\label{eq:multigraph_fusing}
    \hat{\mathds{1}}_{\mathcal{K}_G}(\{E_e\}_{e\in\mathcal{E}})=\hat{\mathds{1}}_{\mathcal{K}_{G'}}(\{E_e\}_{e\in\mathcal{E}'})\big|_{E_e=\sum_{j=1}^n E_j}.
\end{equation}
In plain words, parallel edges are substituted with a unique edge whose energy equals the sum of their energies. This replacement will result in a significant simplification of the combinatorial factors involved in the computation of $\hat{\mathds{1}}_{\mathcal{K}_G}$. This property of the Fourier transform of the cone $\mathcal{K}_G$ mirrors an analogous treatment of multigraphs presented in other works~\cite{Bloch:2015efx,Kreimer:2020mwn,Sborlini_2021}.

We now present the fundamental relation required to triangulate the cone $\mathcal{K}_G$. Let $G$ be a simple digraph and let us consider a cut $S\subset V$. Furthermore, let us impose that $\delta(S)=\delta^{+}(S)$. The existence of such a cut is guaranteed by the acyclic property. 
Then
\begin{equation}
\label{eq:contraction_formula}
        \hat{\mathds{1}}_{\mathcal{K}_G}=\frac{1}{i(\mathbf{E}-\mathbf{p}_G^0)\cdot \boldsymbol{1}^{\delta(S)}}\sum_{a\in\delta(S)}\hat{\mathds{1}}_{\mathcal{K}_{G_a}}.
\end{equation}
where $G_a=(\mathcal{V}_a,\mathcal{E}_a)$ is the graph obtained from $G$ by contracting the edge $a$. Note that $\mathbf{p}_G^0\cdot\mathbf{1}^{\delta(S)}=\sum_{v\in S}p_v^0$ if $\delta^{+}(S)=\delta(S)$. Comparison with general formulas for the Fourier transform of cones~\cite{polytope_fourier} establishes that $\boldsymbol{1}^{\delta(S)}$ is one of the edge vectors emanating from the vertex of the cone. Eq.~\eqref{eq:contraction_formula} is readily derived given that
\begin{equation}
    \prod_{e\in\delta(S)}\Theta(\tau_e)=\sum_{a\in\delta(S)}\Theta(\tau_a)\prod_{e\in\delta(S)\setminus\{a\}}\Theta(\tau_e-\tau_a),
\end{equation}
which expresses the triangulation of the positive orthant $\mathbb{R}_+^{|\delta(S)|}$ in $|\delta(S)|$ cones. Once substituted in $\hat{\mathds{1}}_{\mathcal{K}_G}$, we obtain a sum of $|\delta(S)|$ integrals, labelled by an index $a\in\delta(S)$. We then perform the change of variables $\tau_{e}'=\tau_e-\tau_a$, $e\in\delta(S)\setminus\{a\}$ for the integral labelled by $a$. This factorises integration over $\tau_a$ completely, which is then trivially performed, yielding eq.~\eqref{eq:contraction_formula}. 

\section{Applications}

\subsection{Singularities, connectedness and crossing}
\label{sec:connectedness_crossing}

\begin{figure*}[ht]
\begin{flushleft}
	\input{cross_free_families/mercedes_diagrams/mercedes} \\[-0.3cm]
	\raisebox{0.25cm}{\resizebox{5.17cm}{!}{$\displaystyle{=\frac{-i}{E_1+E_2+E_3+p_1^0}}\frac{-i}{E_3+E_5+E_7-p_3^0}\Bigg[$}}\raisebox{-0.52cm}{\raisebox{0.35cm}{\resizebox{1.7cm}{!}{
\begin{tikzpicture}

    \begin{feynman}
        \vertex(1);

        \vertex[right = 1cm of 1](C);
        
        \vertex[right = 0cm of 1](A1);
        \vertex[right = 1cm of C](A3);
        \vertex[above = 1cm of C](A2);
        \vertex[below = 1cm of C](A4);
        
        \vertex[left = 0.1cm of A1](L1){\resizebox{0.9cm}{!}{$v_{135}$}};
        \vertex[right = 0.1cm of A3](L3){\resizebox{0.5cm}{!}{$v_4$}};
        \vertex[right = 0.1cm of C](L4){\resizebox{0.5cm}{!}{$v_2$}};

         \diagram*[large]{	
            (A1) -- [->,line width=0.5mm, quarter left, blue!80!white] (A2) -- [-,line width=0.5mm, quarter left, blue!80!white] (A3),
            
            (A4) -- [<-,line width=0.5mm, quarter right, blue!80!white] (A3),
            (A4) -- [-,line width=0.5mm, quarter left, blue!80!white] (A1), 
            
            (A1) -- [->-,line width=0.5mm] (C),
            (A1) -- [->-,line width=0.5mm, half left] (C),
            (A1) -- [-<-,line width=0.5mm, half right] (C),


        }; 
       \path[draw=black, fill=black] (A1) circle[radius=0.08];
       \path[draw=black, fill=black] (A3) circle[radius=0.08];
       \path[draw=black, fill=black] (C) circle[radius=0.08];
       
       
    \end{feynman}

\end{tikzpicture}
}}}\raisebox{0.25cm}{$\displaystyle{+}$}
\raisebox{-0.15cm}{\resizebox{1.65cm}{!}{
\begin{tikzpicture}

    \begin{feynman}
        \vertex(1);

        \vertex[right = 1cm of 1](C);
        
        \vertex[right = 0cm of 1](A1);
        \vertex[right = 1cm of C](A3);
        \vertex[above = 1cm of C](A2);
        \vertex[below = 1cm of C](A4);
        
        \vertex[left = 0.1cm of A1](L1){\resizebox{0.75cm}{!}{$v_{15}$}};
        \vertex[right = 0.1cm of A3](L3){\resizebox{0.75cm}{!}{$v_{23}$}};
        \vertex[above = 0.1cm of A2](L2){\resizebox{0.5cm}{!}{$v_4$}};

         \diagram*[large]{	
            (A1) -- [->-,line width=0.5mm, quarter left] (A2) -- [->-,line width=0.5mm, quarter left] (A3),
            
            (A4) -- [-,line width=0.5mm, quarter right] (A3),
            (A4) -- [<-,line width=0.5mm, quarter left] (A1), 
            
            (A1) -- [->-,line width=0.5mm, out=-35, in=-145] (A3),
            (A1) -- [->-,line width=0.5mm, out=35, in=145] (A3),


        }; 
       \path[draw=red, fill=red!80!blue] (A1) circle[radius=0.1];
       \path[draw=black, fill=black] (A2) circle[radius=0.08];
       \path[draw=black, fill=black] (A3) circle[radius=0.08];
       
       
    \end{feynman}

\end{tikzpicture}
}}\raisebox{0.25cm}{$\displaystyle{+}$}
\raisebox{-0.15cm}{\resizebox{1.65cm}{!}{%
\begin{tikzpicture}

    \begin{feynman}
        \vertex(1);

        \vertex[right = 1cm of 1](C);
        
        \vertex[right = 0cm of 1](A1);
        \vertex[right = 1cm of C](A3);
        \vertex[above = 1cm of C](A2);
        \vertex[below = 1cm of C](A4);
        
        \vertex[left = 0.1cm of A1](L1){\resizebox{0.75cm}{!}{$v_{15}$}};
        \vertex[right = 0.1cm of A3](L3){\resizebox{0.75cm}{!}{$v_{34}$}};
        \vertex[above = 0.1cm of C](L4){\resizebox{0.5cm}{!}{$v_2$}};

         \diagram*[large]{	
            (A1) -- [->,line width=0.5mm, quarter left] (A2) -- [-,line width=0.5mm, quarter left] (A3),
            
            (A4) -- [-,line width=0.5mm, quarter right] (A3),
            (A4) -- [<-,line width=0.5mm, quarter left] (A1), 
            
            (A1) -- [->-,line width=0.5mm] (C),
            (A1) -- [->-,line width=0.5mm, half right] (C),
            (C) -- [->-,line width=0.5mm] (A3),


        }; 
       \path[draw=red, fill=red!80!blue] (A1) circle[radius=0.1];
       \path[draw=black, fill=black] (A3) circle[radius=0.08];
       \path[draw=black, fill=black] (C) circle[radius=0.08];
       
       
    \end{feynman}

\end{tikzpicture}
}}\raisebox{0.25cm}{\resizebox{0.17cm}{!}{$\Bigg]$}} \\[-0.1cm]
	\raisebox{0.25cm}{\resizebox{9cm}{!}{$\displaystyle{=\frac{-i}{E_1+E_2+E_3+p_1^0}}\frac{-i}{E_3+E_5+E_7-p_3^0}\frac{-i}{E_2+E_3+E_4+E_6+p_1^0+p_5^0}\Bigg[$}}\hspace{-0.1cm}\raisebox{-0.16cm}{
\resizebox{1.8cm}{!}{%
\begin{tikzpicture}

    \begin{feynman}
        \vertex(1);

        \vertex[right = 0.75cm of 1](C);
        
        
        \vertex[right = 0cm of 1](A1);
        \vertex[right = 0.75cm of C](A3);
        \vertex[above = 0.75cm of C](A2);
        \vertex[below = 0.75cm of C](A4);
        
        \vertex[left = 0.1cm of A1](L1){\resizebox{0.75cm}{!}{$v_{154}$}};
        \vertex[right = 0.1cm of A3](L3){\resizebox{0.6cm}{!}{$v_{23}$}};

         \diagram*[large]{	
            (A1) -- [->,line width=0.4mm, quarter left] (A2) -- [-,line width=0.4mm, quarter left] (A3),
            
            (A4) -- [-,line width=0.4mm, quarter right] (A3),
            (A4) -- [<-,line width=0.4mm, quarter left] (A1), 
            
            (A3) -- [-<-,line width=0.4mm, out=145, in=35] (A1),
            (A3) -- [-<-,line width=0.4mm, out=-145, in=-35] (A1),


        }; 
       \path[draw=red, fill=red!80!blue] (A1) circle[radius=0.1];
       \path[draw=black, fill=black] (A3) circle[radius=0.08];
       
       
    \end{feynman}

\end{tikzpicture}
}}\raisebox{0.25cm}{$+$}
\hspace{-0.1cm}\raisebox{-0.16cm}{
\resizebox{1.8cm}{!}{%
\begin{tikzpicture}

    \begin{feynman}
        \vertex(1);

        \vertex[right = 0.75cm of 1](C);
        
        
        \vertex[right = 0cm of 1](A1);
        \vertex[right = 0.75cm of C](A3);
        \vertex[above = 0.75cm of C](A2);
        \vertex[below = 0.75cm of C](A4);
        
        \vertex[left = 0.1cm of A1](L1){\resizebox{0.95cm}{!}{$v_{1235}$}};
        \vertex[right = 0.1cm of A3](L3){\resizebox{0.4cm}{!}{$v_{4}$}};

         \diagram*[large]{	
            (A1) -- [->,line width=0.5mm, quarter left, blue!80!white] (A2) -- [-,line width=0.5mm, quarter left, blue!80!white] (A3),
            
            (A4) -- [<-,line width=0.5mm, quarter right, blue!80!white] (A3),
            (A4) -- [-,line width=0.5mm, quarter left, blue!80!white] (A1), 
            


        }; 
       \path[draw=black, fill=black] (A1) circle[radius=0.08];
       \path[draw=black, fill=black] (A3) circle[radius=0.08];
       
       
    \end{feynman}

\end{tikzpicture}
}}\raisebox{0.25cm}{$+$}\hspace{-0.1cm}
\raisebox{-0.16cm}{
\resizebox{1.8cm}{!}{%
\begin{tikzpicture}

    \begin{feynman}
        \vertex(1);

        \vertex[right = 0.75cm of 1](C);
        
        
        \vertex[right = 0cm of 1](A1);
        \vertex[right = 0.75cm of C](A3);
        \vertex[above = 0.75cm of C](A2);
        \vertex[below = 0.75cm of C](A4);
        
        \vertex[left = 0.1cm of A1](L1){\resizebox{0.75cm}{!}{$v_{125}$}};
        \vertex[right = 0.1cm of A3](L3){\resizebox{0.65cm}{!}{$v_{34}$}};

         \diagram*[large]{	
            (A1) -- [->,line width=0.4mm, quarter left] (A2) -- [-,line width=0.4mm, quarter left] (A3),
            
            (A4) -- [-,line width=0.4mm, quarter right] (A3),
            (A4) -- [<-,line width=0.4mm, quarter left] (A1), 
            
            (A3) -- [-<-,line width=0.4mm] (A1),


        }; 
       \path[draw=red, fill=red!80!blue] (A1) circle[radius=0.1];
       \path[draw=black, fill=black] (A3) circle[radius=0.08];
       
       
    \end{feynman}

\end{tikzpicture}
}}\raisebox{0.25cm}{$+$}
\hspace{-0.1cm}\raisebox{-0.16cm}{
\resizebox{1.8cm}{!}{%
\begin{tikzpicture}

    \begin{feynman}
        \vertex(1);

        \vertex[right = 0.75cm of 1](C);
        
        
        \vertex[right = 0cm of 1](A1);
        \vertex[right = 0.75cm of C](A3);
        \vertex[above = 0.75cm of C](A2);
        \vertex[below = 0.75cm of C](A4);
        
        \vertex[left = 0.1cm of A1](L1){\resizebox{1cm}{!}{$v_{1534}$}};
        \vertex[right = 0.1cm of A3](L3){\resizebox{0.45cm}{!}{$v_{2}$}};

         \diagram*[large]{	
            (A1) -- [-,line width=0.4mm, quarter left, blue!80!white] (A2) -- [<-,line width=0.4mm, quarter left, blue!80!white] (A3),
            
            (A4) -- [-,line width=0.4mm, quarter right] (A3),
            (A4) -- [<-,line width=0.4mm, quarter left] (A1), 
            
            (A3) -- [-<-,line width=0.4mm, blue!80!white] (A1),


        }; 
       \path[draw=black, fill=black] (A1) circle[radius=0.08];
       \path[draw=black, fill=black] (A3) circle[radius=0.08];
       
       
    \end{feynman}

\end{tikzpicture}
}}\raisebox{0.25cm}{\resizebox{0.17cm}{!}{$\Bigg]$}} \\[0.2cm]
	\input{cross_free_families/mercedes_diagrams/mercedes_4}
\end{flushleft}
	\caption{\label{fig:contraction} Graphical depiction of the recursion on the graph, with edges labelled $e_1=\{v_1,v_5\}, \ e_2=\{v_1,v_2\}, \ e_3=\{v_1,v_3\}, \ e_4=\{v_2,v_5\}, \ e_5=\{v_2,v_3\}, \ e_6=\{v_5,v_4\}, \ e_7=\{v_3,v_4\}$. The cross-free families generated for this acyclic graph are $\mathcal{F}_G=\{F_1,F_2\}$ with $F_1=\{\{v_1\},\{v_3\},\{v_1,v_5\},\{v_{1},v_5,v_4\}\}$ and $F_2=\{\{v_1\},\{v_3\},\{v_1,v_5\},\{v_{1},v_2,v_5\}\}$. At the first step, we choose the source $v_1$, and obtain three graphs obtained by contracting the edges adjacent to $v_1$. The second equality is obtained by observing that two of the graphs have directed cycles (highlighted in blue), so the Fourier transform of their associated cone evaluates to zero. The third equality is obtained by now choosing the sink $v_3$, which yields three graphs, of which two are acyclic. Observe that we contract all parallel edges that connect two vertices. The fourth equality is obtained by choosing the source $v_{15}$ for the two acyclic graphs. Each yields two contracted graphs, for a total of two acyclic graphs. The final equality is obtained by choosing the sources $v_{154}$ and $v_{125}$ for the two acyclic graphs obtained at the previous step. 
}
\end{figure*}

The contraction operation allows to recursively perform all the Fourier integrations defining $\hat{\mathds{1}}_{\mathcal{K}_G}$, a procedure that is equivalent to finding a special triangulation of the convex cone $\mathcal{K}_G$. Such a triangulation is different, for example, than the one needed to obtain the TOPT representation. The specific triangulation we choose makes the relationship between crossing, connectedness and the threshold singularity structure of Feynman diagrams manifest. A large body of recent works~\cite{Abreu:2014cla,Bloch:2015efx,Arkani-Hamed:2017fdk,Abreu:2017ptx,Capatti:2020xjc,bourjaily2020sequential,Berghoff:2020bug,Kreimer:2020mwn,Kreimer:2021jni,Benincasa:2020aoj,Sborlini_2021,Hannesdottir} refers to this correspondence with varying degree of rigour and in separate contexts, which motivates looking for an effective and concise description of it. In particular, we will provide a combinatorial and generic construction procedure for the \emph{Cross-Free Family representation} of Feynman integrals that was conjectured in eq.~(3.44) of~\cite{Capatti:2020xjc}. 

First, given a simple digraph $G$ with vertices labelled as $\mathcal{V}=\{v_{I_1},...,v_{I_n}\}$ for given subsets $I_1,...,I_n\subseteq\mathcal{X}$ of a ground set $\mathcal{X}$ of discrete elements, and given an edge $a=\{v_{I_i},v_{I_j}\}$ to be contracted, we denote with 
\begin{equation}
\label{eq:edge_contraction}
    G_a=(\mathcal{V}\setminus\{v_{I_i},v_{I_j}\}\cup\{v_{I_i\cup I_j}\},\mathcal{E}\setminus \{a\}).
\end{equation} 
the contracted graph. This notation allows to map quantities in the contracted graph to quantities in the original graph. Then given, as an input, the simple, acyclic graph $G_i$ and the family $F_i\subset\mathcal{P}(\mathcal{X})$ ($\mathcal{P}(\mathcal{X})$ being the powerset of $\mathcal{X}$):
\begin{itemize}
\renewcommand\labelitemi{\tiny$\bullet$}
\item Find a source or a sink $v_I$ of the graph $G_i$ such that $(\mathcal{V}_i\setminus\{v_I\}, \mathcal{E}_i\setminus\delta(v_I))$ is a connected graph. The existence of such a vertex is guaranteed by the acyclic property of the graph $G_i$. Let $F_{i+1}=F_i\cup\{I\}$.
\item Let $\{(G_i)_a\}_{a\in\delta(v)}$ be the collection of graphs contracted according to the rule of eq.~\eqref{eq:edge_contraction}.
\item For each of the graphs $(G_i)_a$, if the graph is not acyclic, terminate and output nothing. If the graph $(G_i)_a$ consists of a single vertex, terminate and output $F_{i+1}$. If none of the two applies, fuse all parallel edges, as described in eq.~\eqref{eq:multigraph_fusing}, so that the resulting graph is also simple, and iterate.
\end{itemize}
This algorithmic procedure is guaranteed to end with a graph that consists of a single vertex. The starting graph $G_{\text{init}}$ is defined to have the set of vertices $\mathcal{V}_{\text{init}}=\{v_{\{1\}},...,v_{\{n\}}\}$, and $F_{\text{init}}=\emptyset$. It follows that, at each iteration, the set $I$, as well as any set in $F_i$, can be mapped to a subset of $\mathcal{V}_{\text{init}}$. The output of the algorithm is a collection of families $\mathcal{F}_G$, dependent on the choice of source or sink at each iteration. Each family $F_{\text{out}}\in\mathcal{F}_G$ is a collection of cuts $S\in F_{\text{out}}$, themselves collections of vertices $S\subset\mathcal{V}_{\text{init}}$.

The families of cuts $F_{\text{out}}=\{S_1,...,S_{|V|-1}\}$ obtained in such a way satisfy the following properties:
\begin{enumerate}
    \item[\small a1)] $S_i$ is connected
    \item[\small a2)] $V\setminus S_i$ is connected
    \item[\small b)] $F_{\text{out}}$ is a laminar family, that is for any two sets $S,S'\in F_{\text{out}}$ of the family, either $S$ and $S'$ are contained one in the other, or $S\cap S'=\emptyset$.
    \item[\small c)] $F_{\text{out}}$ is obstruction-free, that is no element of the family can be written as a union of sets contained within it.
\end{enumerate}
Given the set $\mathcal{F}_G$, we are able to evaluate the Fourier transform of the cone $\mathcal{K}_G$. The algorithm presented above effectively performs diagrammatically the triangulation of the cone $\mathcal{K}_{G}$ in $|\mathcal{F}_{G}|$ simplicial cones, with each \textit{cross-free family} $F\in\mathcal{F}_{G}$ representing a simplicial cone. Using it, one can evaluate all Fourier integrations and obtain the CFF representation
\begin{equation}
    \colorboxed{green!80!blue}{ \, I_{G_u}=\sum_{G\in\text{dag}(G_u)}\sum_{F\in\mathcal{F}_G} \frac{i^L(\prod_{e\in \mathcal{E}}2E_e)^{-1}\mathcal{N}_G(\mathbf{E})}{\prod_{S\in F} (\mathbf{p}_G^0-\mathbf{E})\cdot \boldsymbol{1}^{\delta(S)}},\,  }
\end{equation}
which provides a proof by construction of the representation conjectured in~\cite{Capatti:2020xjc}. We provide in the mathematica package \texttt{cLTD.m}~\footnote{ \href{http://www.github.com/apelloni/cLTD}{github.com\slash apelloni\slash cLTD}} a generic implementation of this algorithmic procedure, resulting in a ready-to-evaluate integrand. 

\subsection{Diagram-level factorisation and iterated connectedness}
\label{sec:diagram_factorisation}

 Factorisation formulas for Fourier transforms of cones and polytopes can be used to effectively study the leading behaviour of the integrand in singular limits~\cite{Borinsky:2022msp}. In particular, the Fourier transform $\hat{\mathds{1}}_{\mathcal{K}_G}$ satisfies a factorisation formula. Let us consider a \emph{connected} cut $S$ with \emph{connected} complement and such that $\delta(S)=\delta^{+}(S)$, let $(\mathbf{E}-\mathbf{p}^0_G)\cdot \boldsymbol{1}^{\delta(S)}=\epsilon$ and consider the leading contribution to $\hat{\mathds{1}}_{\mathcal{K}_G}(\{p_v^0\}_{v\in\mathcal{V}})$ in the expansion in $\epsilon$, which is
\begin{equation}
\label{eq:diagram_factorisation}
    \hat{\mathds{1}}_{\mathcal{K}_G}=\frac{\hat{\mathds{1}}_{\mathcal{K}_{G_1}}(\{p_v^{G_1}\}_{v\in\mathcal{V}_1}) \hat{\mathds{1}}_{\mathcal{K}_{G_2}}(\{p_v^{G_2}\}_{v\in\mathcal{V}_2})}{\epsilon} +o(1).
\end{equation}
$p_v^{G_i}$, $v\in\mathcal{V}_i$, $i=1,2$, are the capacities for the graphs $G_1=(\mathcal{V}_1=S,\mathcal{E}_1)$ and $G_2=(\mathcal{V}_2=\mathcal{V}\setminus S,\mathcal{E}_2)$, obtained from $G$ by deleting the edges in $\delta(S)$. The external energies for the two graphs are defined as follows: if, for a given edge $e$, we let $v_{e}^{(1)}$ be its departing vertex and $v_{e}^{(2)}$ its arriving one, then
\begin{equation}
    p^{G_i}_v=\begin{cases}
    p_v^0+(-1)^i (E_e-p_e^0) \quad \text{if } v=v_{e}^{(i)}, \ e\in\delta(S) \\
    p_v^0 \quad \text{otherwise}
    \end{cases}.
\end{equation}
Eq.~\eqref{eq:diagram_factorisation} is readily obtained by direct evaluation of the integrals in the variables $\tau_e$, $e\in\delta(S)$ at leading order in $\epsilon$. Importantly, if the cut $S$ is disconnected or has disconnected complement, then $\mathds{1}_{\mathcal{K}_G}=o(1)$.  

Given eq.~\eqref{eq:diagram_factorisation}, we now iterate the argument. The singularities of the two graphs $G_1$ and $G_2$ correspond to connected cuts such that the complement is also connected. Furthermore, these singularities must correspond to singularities of the original graph $G$, evaluated at $(\mathbf{E}-\mathbf{p}^0_G)\cdot\mathbf{1}^{\delta(S)}=0$. Thus, let us consider a cross-free family $F$ of cuts, and let $(\mathbf{E}-\mathbf{p}^0_G)\cdot\mathbf{1}^{\delta(S)}=o(\epsilon)$, for all $S\in F$, and $(\mathbf{E}-\mathbf{p}^0_G)\cdot\mathbf{1}^{\delta(S)}=o(1)$ for $S\notin F$. By iterating the factorisation argument, we obtain that a necessary and sufficient condition for
\begin{equation}
\label{eq:iterated_connectedness}
    \hat{\mathds{1}}_{\mathcal{K}_G}(\mathbf{E},\{p_v^0\}_{v\in\mathcal{V}})=\frac{w(\mathbf{E},\{p_v^0\}_{v\in\mathcal{V}})}{\epsilon^{|F|}}+o(\epsilon^{-|F|+1})
\end{equation}
to hold with a non-vanishing function $w$ is that the cuts in $F$ divide the graph in $|F|+1$ connected components, which is the lowest possible value. We then say that $F$ satisfies the \emph{iterated connectedness} property. 

\begin{figure}[t]

\begin{center}
	\resizebox{8cm}{!}{
\raisebox{-0.27cm}{\resizebox{2.7cm}{!}{%
\begin{tikzpicture}

    \begin{feynman}
        \vertex(1);

        \vertex[above = 1cm of 1](A1);
        
        \vertex[above right = 0.309016994375cm and 0.951056516295cm of 1](A2);
        \vertex[below right = 0.809016994375cm and 0.587785252292cm of 1](A3);
        \vertex[below left = 0.809016994375cm and 0.587785252292cm of 1](A4);
        \vertex[above left = 0.309016994375 and 0.951056516295 of 1](A5);
        
        \vertex[above = 0.1cm of A1](L1){\resizebox{0.5cm}{!}{$v_1$}};
        \vertex[right = 0.1cm of A2](L3){\resizebox{0.5cm}{!}{$v_2$}};
        \vertex[left = 0.1cm of A5](L2){\resizebox{0.5cm}{!}{$v_5$}};
        \vertex[below left = 0.05cm and 0.05cm of A4](L4){\resizebox{0.5cm}{!}{$v_4$}};
        \vertex[below right = 0.05cm and 0.05cm of A3](L3){\resizebox{0.5cm}{!}{$v_3$}};

         \diagram*[large]{	
            (A1) -- [->-,line width=0.5mm] (A5),
            (A2) -- [->-,line width=0.5mm] (A1),
            (A4) -- [->-,line width=0.5mm] (A5),
            (A3) -- [->-,line width=0.5mm] (A4),
            (A2) -- [->-,line width=0.5mm] (A3),

        }; 
       \path[draw=black, fill=black] (A1) circle[radius=0.05];
       \path[draw=black, fill=black] (A2) circle[radius=0.05];
       \path[draw=black, fill=black] (A3) circle[radius=0.05];
       \path[draw=black, fill=black] (A4) circle[radius=0.05];
       \path[draw=black, fill=black] (A5) circle[radius=0.05];
       
       
    \end{feynman}

\end{tikzpicture}
}}\hspace{0.cm}\raisebox{0.6cm}{$=$}
\raisebox{-0.cm}{\resizebox{2.cm}{!}{%
\begin{tikzpicture}

    \begin{feynman}
        \vertex(1);

        \vertex[above = 1cm of 1](A1);
        
        \vertex[above right = 0.309016994375cm and 0.951056516295cm of 1](A2);
        \vertex[below right = 0.809016994375cm and 0.587785252292cm of 1](A3);
        \vertex[below left = 0.809016994375cm and 0.587785252292cm of 1](A4);
        \vertex[above left = 0.309016994375 and 0.951056516295 of 1](A5);
        

         \diagram*[large]{	
            (A1) -- [->-,line width=0.5mm] (A5),
            (A2) -- [->-,line width=0.5mm] (A1),
            (A4) -- [->-,line width=0.5mm] (A5),
            (A3) -- [->-,line width=0.5mm] (A4),
            (A2) -- [->-,line width=0.5mm] (A3),

        }; 
       \path[draw=black, fill=black] (A1) circle[radius=0.05];
       \path[draw=black, fill=black] (A2) circle[radius=0.05];
       \path[draw=black, fill=black] (A3) circle[radius=0.05];
       \path[draw=black, fill=black] (A4) circle[radius=0.05];
       \path[draw=black, fill=black] (A5) circle[radius=0.05];
       
    \end{feynman}
    
    \draw[rotate=45,line width=0.25mm,magenta!90!blue] (A5) ellipse (5pt and 5pt);
    \draw[rotate=35,line width=0.25mm,magenta!90!blue] (-0.05,0.8) ellipse (24pt and 11pt);
    \draw[rotate=70,line width=0.25mm,magenta!90!blue] (0.05,0.5) ellipse (35pt and 25pt);
    \draw[rotate=45,line width=0.25mm,magenta!90!blue] (A2) ellipse (5pt and 5pt);

\end{tikzpicture}
}}\hspace{0.0cm}\raisebox{0.6cm}{$+$}
\raisebox{-0.08cm}{\resizebox{1.93cm}{!}{%
\begin{tikzpicture}

    \begin{feynman}
        \vertex(1);

        \vertex[above = 1cm of 1](A1);
        
        \vertex[above right = 0.309016994375cm and 0.951056516295cm of 1](A2);
        \vertex[below right = 0.809016994375cm and 0.587785252292cm of 1](A3);
        \vertex[below left = 0.809016994375cm and 0.587785252292cm of 1](A4);
        \vertex[above left = 0.309016994375 and 0.951056516295 of 1](A5);
        

         \diagram*[large]{	
            (A1) -- [->-,line width=0.5mm] (A5),
            (A2) -- [->-,line width=0.5mm] (A1),
            (A4) -- [->-,line width=0.5mm] (A5),
            (A3) -- [->-,line width=0.5mm] (A4),
            (A2) -- [->-,line width=0.5mm] (A3),

        }; 
       \path[draw=black, fill=black] (A1) circle[radius=0.05];
       \path[draw=black, fill=black] (A2) circle[radius=0.05];
       \path[draw=black, fill=black] (A3) circle[radius=0.05];
       \path[draw=black, fill=black] (A4) circle[radius=0.05];
       \path[draw=black, fill=black] (A5) circle[radius=0.05];
       
    \end{feynman}
    
    \draw[rotate=45,line width=0.25mm,magenta!90!blue] (A5) ellipse (5pt and 5pt);
    \draw[rotate=105,line width=0.25mm,magenta!90!blue] (-0.0,0.82) ellipse (24pt and 11pt);
    \draw[rotate=45,line width=0.25mm,magenta!90!blue] (A2) ellipse (5pt and 5pt);
    \draw[rotate=55,line width=0.25mm,magenta!90!blue] (0.8,-0.06) ellipse (11pt and 24pt);

\end{tikzpicture}
}}\hspace{0.0cm}\raisebox{0.6cm}{$+$}
\raisebox{-0.22cm}{\resizebox{1.97cm}{!}{%
\begin{tikzpicture}

    \begin{feynman}
        \vertex(1);

        \vertex[above = 1cm of 1](A1);
        
        \vertex[above right = 0.309016994375cm and 0.951056516295cm of 1](A2);
        \vertex[below right = 0.809016994375cm and 0.587785252292cm of 1](A3);
        \vertex[below left = 0.809016994375cm and 0.587785252292cm of 1](A4);
        \vertex[above left = 0.309016994375 and 0.951056516295 of 1](A5);
        

         \diagram*[large]{	
            (A1) -- [->-,line width=0.5mm] (A5),
            (A2) -- [->-,line width=0.5mm] (A1),
            (A4) -- [->-,line width=0.5mm] (A5),
            (A3) -- [->-,line width=0.5mm] (A4),
            (A2) -- [->-,line width=0.5mm] (A3),

        }; 
      \path[draw=black, fill=black] (A1) circle[radius=0.05];
       \path[draw=black, fill=black] (A2) circle[radius=0.05];
       \path[draw=black, fill=black] (A3) circle[radius=0.05];
       \path[draw=black, fill=black] (A4) circle[radius=0.05];
       \path[draw=black, fill=black] (A5) circle[radius=0.05];
       
    \end{feynman}
    
    \draw[rotate=45,line width=0.25mm,magenta!90!blue] (A5) ellipse (5pt and 5pt);
    \draw[rotate=75,line width=0.25mm,magenta!90!blue] (-0.02,-0.83) ellipse (24pt and 11pt);
    \draw[rotate=45,line width=0.25mm,magenta!90!blue] (A2) ellipse (5pt and 5pt);
    \draw[rotate=-68,line width=0.25mm,magenta!90!blue] (0.13,0.48) ellipse (35pt and 25pt);

\end{tikzpicture}
}}
} 
\end{center}
	\caption{\label{fig:pentagon} A diagrammatic way to represent the result of performing by contraction the integrals of $\mathds{1}_{\mathcal{K}_G}$ is to draw the resulting cross-free families. The three diagrams that are obtained as a result of the triangulation correspond to the cross-free families $F_1=\{\{v_5\},\{v_1,v_5\},\{v_1,v_4,v_5\},\{v_2\}\}$, $F_2=\{\{v_5\},\{v_4,v_5\},\{v_1,v_2\},\{v_2\}\}$, $F_3=\{\{v_5\},\{v_2\},\{v_2,v_3\},$ $\{v_1,v_2,v_3\}\}$, respectively. Each of the three terms scales like $\epsilon^{-2}$ in the limit in which the two thresholds $\{v_5\}$ and $\{v_2\}$ vanish. However, their sum scales like $\epsilon^{-1}$ since they divide the graph in four connected components instead of three, in agreement with the bound of sect.~\ref{sec:diagram_factorisation}.
}
\end{figure}
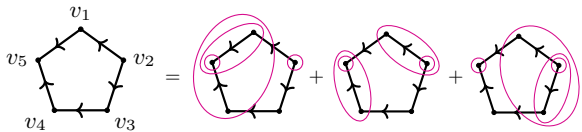

\section{Outlook}

Let us summarise our findings and their implications. In sect.~\ref{sec:connectedness_crossing} we found a new three-dimensional representation of Feynman diagrams, called the \emph{cross-free family representation}, that is free of spurious singularities and in which the relation between thresholds and connected subgraphs as well as the relation between intersections of thresholds and crossing of subgraphs are manifest. In sect.~\ref{sec:diagram_factorisation}, we derived an upper bound on the scaling of three-dimensional representations close to the intersection of a number of thresholds. When combined with a simple scaling argument, the bound determines that if $n$ thresholds of a QCD diagram intersect at a point, then the singularity at that point is non-integrable only if the corresponding cuts divide the graph in $n+1$ connected components. This principle may be used to drastically simplify regularisation procedures of pinched and non-pinched thresholds~\cite{Anastasiou:2018rib,Anastasiou:2020sdt,Ma:2019hjq,Capatti_2020,Kermanschah:2021wbk,Becker:2010ng,Becker:2012aqa,Becker:2012nk,Gong:2008ww,Buchta:2015wna,Soper:1999xk}.

We conclude by the discussing the deep implications of the factorisation formula of sect.~\ref{sec:diagram_factorisation} on the KLN-based cancellation pattern~\cite{Kinoshita:1962ur,Lee:1964is,Frye:2018xjj,Hannesdottir:2019rqq,Capatti:2020xjc,Hannesdottir:2022bmo,Khalil:2017fbz,Capatti:2022tit,Khalil:2017yiy,Blazek:2021zoj,Lavelle_2006,Akhoury_1997,Sborlini:2016gbr,Gomez:2018war,Carney:2017jut,Melnikov_1997,DILIETO1981223,FRAZER1958137,CARNEIRO1981445,10.1143/ptp/5.6.1045,PhysRevD.25.2222,PhysRevD.32.2385} of infrared singularities. One important consequence is the following: consider the quantity
\begin{equation}
    P=\sum_{n,m=0}^\infty a_{nm} \mathrm{Tr}[\, \hat{\boldsymbol{\mathrm{\rho}}} \, \hat{\mathbf{S}}_\mathrm{c}^n \, \hat{\boldsymbol{\mathrm{P}}} \, (\hat{\mathbf{S}}_\mathrm{c}^\dagger)^m],
\end{equation}
where $\hat{\boldsymbol{\mathrm{\rho}}}$ and $\hat{\boldsymbol{\mathrm{P}}}$ are matrices in the Fock space of the free theory that describe the density of initial states and measurement respectively. They both are diagonal in the momenta and quantum numbers of the single particle states. $\hat{\mathbf{S}}_\mathrm{c}$ is the connected scattering matrix which schematically takes the form

\begin{equation}\label{eq:diag_rep_s_matrix}
\bra{\boldsymbol{\alpha}}\hat{\mathbf{S}}_\mathrm{c} \ket{\boldsymbol{\beta}} \, \, = \, \,
\resizebox{1.9cm}{!}{%
\begin{tikzpicture}[baseline=-8ex]
    
    \node[main node2]  (1) {$ \resizebox{1cm}{!}{$G_{\mathrm{c}}$} $};

    \node[] (p2) [below right= 0.8cm and 1.5cm of 1] {};
    \node[] (p1) [below left= 0.8cm and 1.5cm of 1] {};
    
    \node[] (c1) [below = 0.2cm of p1] {};
    \node[] (c2) [below = 0.2cm of p2] {};
    
    \node[] (d1) [below = 0.4cm of c1] {};
    \node[] (d2) [below = 0.4cm of c2] {};
    
    \node[] (label) [below = 0.25cm of 1] {\resizebox{0.2cm}{!}{$\boldsymbol{\vdots}$}};

    \node[] (att1) [above left=-1.5cm and 0.9cm of 1] {\resizebox{0.2cm}{!}{$\boldsymbol{\vdots}$}};
    \node[] (att1p) [above right=-1.5cm and 0.9cm of 1]{\resizebox{0.2cm}{!}{$\boldsymbol{\vdots}$}};
    \node[] (att2) [above left= -0.5cm and 1.5cm of 1] {};
    \node[] (att3) [below left= -0.5cm and 1.5cm of 1] {};

    \node[] (btt2) [above right= -0.5cm and 1.5cm of 1] {};
    \node[] (btt3) [below right= -0.5cm and 1.5cm of 1] {};

    \node[] (cutleft1) [above left= 0.4cm and 0.60cm of 1] {};
    \node[] (cutleft2) [below left= 2.4cm and 0.35cm of 1] {\resizebox{0.5cm}{!}{$\boldsymbol{\alpha}$}};
    
    \node[] (cutright1) [above right= 0.4cm and 0.60cm of 1] {};
    \node[] (cutright2) [below right= 2.4cm and 0.35cm of 1] {\resizebox{0.5cm}{!}{$\boldsymbol{\beta}$}};

	\draw[->,line width=0.5mm] (p1) to (p2); 
	\draw[->,line width=0.5mm] (d1) to (d2); 
	
	\draw[->,line width=0.5mm] (1) to[out= 20, in=180] (btt2);
	\draw[->,line width=0.5mm] (1) to[out= -20, in=180] (btt3);
	
	\draw[->,line width=0.5mm] (att2) to[in= 160, out=0] (1);
	\draw[->,line width=0.5mm] (att3) to[in= -160, out=0] (1);
	
	\draw[-,line width=0.6mm,dotted,white!40!black] (cutleft1) to(cutleft2);
	
	\draw[-,line width=0.6mm,dotted,white!40!black] (cutright1) to (cutright2);

\end{tikzpicture}
}.
\end{equation}
with an implicit sum over all possible ways to partition $\boldsymbol{\alpha}$ and $\boldsymbol{\beta}$ into spectator and scattering states.
$P$ is IR-finite for any value of the coefficients $a_{nm}$, given IR-safety contraints on $\hat{\boldsymbol{\mathrm{\rho}}}$ and $\hat{\boldsymbol{\mathrm{P}}}$. This agrees with the principle implied by eq.~\eqref{eq:iterated_connectedness}, that KLN cancellations happen between interference diagrams that, after the deletion of the cut edges, have the same amount of connected components.

\section{Acknowledgments}

I would like to thank Michael Borinsky, Eric Laenen and Alex Salas-Bern\'adez as my collaboration with them inspired me with some of the proof techniques used in this paper. I am also grateful to Valentin Hirschi and Ben Ruijl for countless discussions on the topic, and for their help with the testing of the cross-free family representation as well as with its implementation. I would also like to thank Babis Anastasiou, Hofie Hannesdottir, Dario Kermanschah and Andrea Pelloni for interesting exchanges on the topic, and Michael Borinsky, Valentin Hirschi, Lucien Huber and Ben Ruijl for their comments on a draft of this paper. This project has received funding from ETH Z\"urich under grant agreement ETH-53 19-2.



\nocite{*}

\bibliographystyle{apsrev4-1}
\bibliography{biblio}

\end{document}